# Are Educational Escape Rooms More Effective Than Traditional Lectures for Teaching Software Engineering? A Randomized Controlled Trial

Aldo Gordillo[ID] and Daniel López-Fernández[ID]

*Abstract*—*Contribution:* This article analyzes the learning effectiveness of a virtual educational escape room for teaching software engineering and compares this activity with traditional teaching through a randomized controlled trial.

*Background:* Educational escape rooms have been used across a wide variety of disciplines at all levels of education and they are becoming increasingly popular among teachers. Nevertheless, there is a clear general need for more robust empirical evidence on the learning effectiveness of these novel activities and, particularly, on their application in software engineering education.

*Research Questions:* Is game-based learning using educational escape rooms more effective than traditional lectures for teaching software engineering? What are the perceptions of software engineering students toward game-based learning using educational escape rooms?

*Methodology:* The study presented in this article is a randomized controlled trial with a pre- and post-test design that was completed by a total of 326 software engineering students. The 164 students belonging to the experimental group learned software modeling by playing an educational escape room whereas the 162 students belonging to the control group learned the same subject matter through a traditional lecture.

*Findings:* The results of the randomized controlled trial show that the students who learned software modeling through the educational escape room had very positive perceptions toward this activity, significantly increased their knowledge, and outperformed those students who learned through a traditional lecture in terms of knowledge acquisition.

## I. INTRODUCTION

OVER the past decade, game-based learning has been applied across a wide range of disciplines in higher education, including software engineering. A recent review of digital game-based learning in engineering education [1], which covered literature from 2011 to 2021 and identified 46 educational games (12 of which were specifically designed for software engineering topics), found that there is substantial evidence suggesting that engineering students can successfully learn by playing games. However, the authors of this review claimed that further investigation is needed on the effectiveness of the game-based learning methodology for engineering education due to the limited number of empirical studies found, the characteristics of the study designs, and the small sample sizes. Specifically, these authors claimed that, in order to make more reliable claims on this matter, more randomized experiments with reasonable sample sizes are necessary.

The aforementioned lack of research on the effectiveness of game-based learning is especially pronounced in the area of educational escape rooms [2], [3], [4], [5], [6], mainly due to the novelty of this kind of learning activities. Educational escape rooms are live-action games where the participating students, usually organized in teams, have to solve a sequence of puzzles within a specified time frame by discovering clues, accomplishing tasks, and applying field-specific knowledge and skills.

Research on educational escape rooms is still in its infancy. A proof of this fact is that the number of articles published in this area before 2017 is practically negligible compared to the current body of knowledge. Although the rise of educational escape rooms has only recently taken place, a total of five systematic literature reviews on the use of escape rooms in education have already been published [2], [3], [4], [5], [6]. These reviews, which have collectively analyzed more than 180 different articles, provide evidence that educational escape rooms have been used across all education levels and disciplines and that they are becoming increasingly popular among teachers. Nevertheless, all of these reviews have concluded that there is currently a lack of empirical evidence on the impact of this type of learning activity and hence further research is needed.

The first literature review on educational escape rooms was published by Fotaris and Mastoras [2] in 2019. This work systematically reviewed 68 studies and found that most of them did not have a control group and resorted to perception surveys to assess how the escape room impacted students' learning and that, even in those cases where knowledge tests or control groups were employed, no trustworthy conclusions could be drawn due to small samples sizes. Another systematic literature review on educational escape rooms was published one year later by Veldkamp et al. [3]. This new review covered

Manuscript received 7 July 2023; revised 6 April 2024; accepted 14 May 2024. This work was supported by the Educational Innovation Project IE22.6105 funded by the Universidad Politécnica de Madrid. *(Corresponding author: Aldo Gordillo.)*

This work involved human subjects or animals in its research. The authors confirm that all human/animal subject research procedures and protocols are exempt from review board approval.

The authors are with the Computer Science Department, Universidad Politécnica de Madrid, 28040 Madrid, Spain (e-mail: a.gordillo@upm.es; daniel.lopez@upm.es).

Digital Object Identifier 10.1109/TE.2024.3403913

39 studies and found that only 3 of them evaluated learning by using pre-post test designs. Furthermore, a discrepancy between self-perceived and objectively measured learning effectiveness of educational escape rooms was discovered.

Makri et al. [4] published in 2021 a systematic literature review centered on virtual educational escape rooms that examined 45 studies and reported that only 4 of them employed pre- and post-tests to objectively measure acquisition of knowledge. Based on these facts, they concluded that more attention should be given to assess the learning outcomes of students engaged in educational escape rooms and specifically suggested conducting more studies employing pretests and post-tests to perform this evaluation.

In another systematic literature review published in 2021, Lathwesen and Belova [5] analyzed 93 studies on educational escape rooms aimed at STEM education and found that most of these studies describe specific escape room experiences with scant or no evidence on their learning effectiveness. In fact, they reported that only 2 out of these 93 studies used a control and experimental group design. These researchers concluded that currently there is no clear evidence on the learning effectiveness of educational escape rooms compared to traditional learning approaches and that there is a general necessity for additional empirical evidence on this issue.

Taraldsen et al. [6] conducted another literature review in which they analyzed 70 articles published between 2017 and 2020 on the utilization of escape rooms within educational settings and classified these articles according to the methods used for evaluating the escape room experiences. In the same vein as the aforementioned literature reviews, the authors of this one concluded that more investigation on the educational application of escape rooms is needed.

In view of the conclusions drawn by the systematic literature reviews on the use of escape rooms in education carried out so far [2], [3], [4], [5], [6], it becomes clear that, although the existing evidence suggests that educational escape rooms hold promise for enhancing student motivation and performance, more research is needed to provide solid empirical evidence on the impact of these novel activities. Indeed, a closer look to these literature reviews reveals that only one of the works analyzed by them has reported a comparison between educational escape rooms and traditional face-to-face teaching through a randomized controlled trial, the most reliable experimental design method for comparing interventions. That work [7] analyzed the utilization of an educational escape room for instructing secondary education students in mathematics and found that the students who learned through this activity outperformed those who learned through traditional teaching in terms of motivation and self-perceived learning effectiveness. In this regard, it should be noted that this work did not employ a pretest post-test design to objectively measure knowledge acquisition and that the activity was not related to software engineering.

Among all the studies analyzed by the aforementioned literature reviews, only eight have reported the application of an educational escape room in a higher-education setting for teaching topics closely related to software engineering [8], [9], [10], [11], [12], [13], [14], [15]. In this aspect, it is noteworthy to mention that none of these studies have performed a comparison between the educational escape room and a traditional approach through a randomized controlled trial or a quasi-experiment using control and experimental groups.

López-Pernas et al. [15] examined an educational escape room for teaching programming in a higher-education course covering front-end web development by collecting the students' perceptions through a survey. In a subsequent work [13], the same authors analyzed a different educational escape room for teaching web programming by using a questionnaire to collect students' perceptions and pre- and post-tests to measure the acquired knowledge. The results demonstrate the successful impact of the escape room in augmenting the students' knowledge. However, given that only 28 students were involved in this study, its findings should be interpreted cautiously due to the small sample size. The same evaluation instruments (i.e., questionnaire, pretests, and post-tests) were used in another study [14] to analyze an educational escape room that was run remotely for teaching software modeling to software engineering students. This study involved 162 students and proved that the escape room was effective and very engaging. Although in this case the sample had a reasonable size, no comparison with a traditional approach was performed.

Otemaier et al. [11] employed a questionnaire to gather the perceptions of software engineering students toward an educational escape room aimed at teaching logical reasoning. No objective measures of acquired knowledge were obtained in this study. Saltz and Heckman [12] employed several educational escape rooms in a data science course for teaching R programming, machine learning, and other topics related to data science. They analyzed the impact of using structured pairing in these escape rooms employing a quasi-experiment with control and experimental groups, but the learning effectiveness was not assessed.

Lastly, three studies have reported the application of educational escape rooms for instructing undergraduate students in cybersecurity topics [8], [9], [10]. The first of these studies presented an educational escape room for teaching cryptographic systems, compression algorithms, and Internet protocols [8]. This escape room was run with two teams of students, but no evaluation was carried out. The second of these studies [9] presented an educational escape room for teaching classic cryptographic systems that required students to use a mathematical software system. This escape room was evaluated merely through student written reflections. The third study [10] describes how an educational escape room was used for teaching cryptographic methods and hash algorithms in a course on software engineering for safe and secure systems. Although that work outlines the assessment of the escape room, no result is reported.

In conclusion, there is a clear need for more robust empirical evidence on the actual learning effectiveness of educational escape rooms at all education levels and disciplines and, particularly, in software engineering education. This article presents a randomized controlled trial with a pre- and post-test design that analyzed the learning effectiveness of an educational escape room for teaching software engineering and compared that activity with traditional teaching. By addressing the aforementioned gap in the literature, this article contributes to a better understanding of the benefits of game-based learning through the use of educational escape rooms and helps educators to gain valuable insights into the effectiveness of these activities as learning tools and make informed decisions about integrating them into their teaching practice.

This article is divided into the following sections. The next section details the research methodology, including a description of the study design, the sample, the procedure, the instruments, the data analysis methods, and the educational escape room examined. The results of the study are reported in Section III whereas the discussion of these results is included in Section IV. This article concludes with Section V, where the conclusions drawn from the study are provided and potential directions for future research are suggested.

## II. RESEARCH METHODOLOGY

### A. Study Design

The study design is a randomized controlled trial with pre and post measurements. The participating students were randomly divided into an experimental group and a control group by using all classrooms of the same course as the unit of randomization. Thus, the study design can be considered a cluster randomized controlled trial, a type of randomized controlled trial in which groups of participants (instead of individual subjects) are the unit of randomization. The students belonging to the experimental group learned software modeling by playing an educational escape room whereas the students belonging to the control group learned the same subject matter through a traditional lecture. Both activities (i.e., the escape room and the traditional lecture) were carried out face-to-face, had the same duration (1 h and a half) and covered the same topics: UML use case diagrams, UML class diagrams, UML sequence diagrams, UML activity diagrams, and UML state machine diagrams.

The research questions addressed in this study are the following.
1) *RQ1:* Is game-based learning using educational escape rooms more effective than traditional lectures for teaching software engineering?
2) *RQ2:* What are the perceptions of software engineering students toward game-based learning using educational escape rooms?

### B. Sample

A total of 326 students completed the randomized controlled trial. Of these students, 164 were part of the experimental group and 162 were part of the control group. All these students were enrolled in a second-year course entitled "Software engineering fundamentals" at Universidad Politécnica de Madrid.

This course is assigned 9 credits according to the European Credit Transfer and Accumulation System (ECTS), which means that each student is expected to complete between 225 and 270 h of work to successfully complete the course.

The course is comprised of five units that cover the following topics, respectively: software processes, requirements analysis and specification, software modeling and analysis, software design, and software testing. According to the curriculum guidelines for undergraduate degree programs in software engineering developed by ACM and the IEEE Computer Society [16], all these topics are essential because they belong to the body of knowledge that every software engineering graduate must know.

### C. Procedure

The randomized controlled trial was carried out halfway through the software engineering fundamentals course previously described. Specifically, it was carried out after finishing the software modeling and analysis unit of the course. In relation to this matter, it is noteworthy to mention that the escape room and the traditional lecture aimed to reinforce the topics related to software modeling that students had already learned in the course. These topics include UML use case diagrams, UML class diagrams, UML sequence diagrams, UML activity diagrams, and UML state machine diagrams.

All participating students gave informed consent to participate and completed all steps of the procedure in person and sequentially. First, each student was randomly assigned to the control or the experimental group as explained in Section II-A. Then, every student took the pretest individually, with a time limit of 10 min for completion.

After this time limit expired, the students carried out the activity corresponding to the group to which they were assigned: the students in the control group received a traditional lecture whereas those in the experimental group played an educational escape room. The traditional lecture lasted 1 h and a half and the same amount of time was set as the time limit to complete the educational escape room. In the traditional lecture conducted in the control group, a course teacher made an oral presentation reviewing the topics related to software modeling previously mentioned and solved a series of test exercises. Additionally, the teacher addressed all doubts raised by the students during the class. For their part, the students in the experimental group participated in the educational escape room in teams of 4–6 members. A brief description of the educational escape room is provided in Section II-F.

Immediately after completing the corresponding activity, all students took the post-test individually, with a maximum completion time of 10 min. In the control group, the post-test was administered immediately after the end of the traditional lecture, while in the experimental group the students took this test either after completing the educational escape room or when the time limit expired. Finally, the students in the experimental group completed the perceptions questionnaire after completing the post-test.

### D. Instruments

Two different types of instruments were used in this randomized controlled trial for collecting participants' data.

First, pre- and post-tests were administered to objectively measure the students' knowledge on software modeling before and after the intervention (i.e., before and after the traditional lecture and the educational escape room). Second, a questionnaire was employed in order to collect the perceptions of the students who played the educational escape room.

The same test was used as pretest and post-test. This test consisted of 10 multiple-choice questions with single answer, was scored on a scale from 0 to 10, and covered the same topics as the traditional lecture and the educational escape room. Thereby, this test allowed to measure the participants' knowledge related to the understanding, interpretation and creation of the following types of UML diagrams: use case, class, sequence, activity, and state machine. Regarding the administration of the tests, it is important to note that no feedback was given to the participating students after finishing the pretest as a measure to prevent them from answering the post-test simply by using their memory.

The perceptions questionnaire included two closed-ended items aimed at gathering the participants' overall opinion about the educational escape room and the web platform used to manage it, two additional closed-ended items related to the acceptance of the activity, and 12 Likert items with responses ranging from 1 (strongly disagree) to 5 (strongly agree). The internal consistency of the perceptions questionnaire was assessed by calculating Cronbach's $\alpha$ [17], which was found to be 0.80, indicating a good level of agreement between the items. Furthermore, the content validity of the perceptions questionnaire was checked by subject matter experts.

### E. Data Analysis

The comparison between post-test and pretest scores within each group was performed by using the Wilcoxon signed-ranks test for paired samples. The comparison of post-test scores between groups was performed by employing ANCOVA with the pretest scores as covariate, whereas pretest scores between groups were compared by employing the Mann–Whitney $U$ test. To measure the effect size in all these comparisons, the correlation coefficient ($r$) was calculated and interpreted in accordance with Cohen's guidelines [18]: small when $0.1 \leq r < 0.3$; medium when $0.3 \leq r < 0.5$; and large when $r \geq 0.5$.

A significance level of 0.05 was used for judging the significance of the differences in all comparisons. The learning gain experienced by each student as a result of receiving the traditional lecture or playing the educational escape room was calculated by subtracting their pretest score from their post-test score. Moreover, pretest scores, post-test scores, learning gains, and Likert items were analyzed by using the mean ($M$) and standard deviation (SD).

### F. Educational Escape Room

The educational escape room was entirely virtual and was conducted face-to-face by means of a web platform called Escapp [19]. This platform allowed students to form teams, access the escape room content once the activity started, view the countdown, the team progress and the leaderboard in real time, verify puzzle solutions, and request hints. The escape room was held in a classroom where each participating student had access to a computer with Internet connection and a web browser with HTML5 support. No other resources were necessary for them to complete the activity. The students played the escape room in teams of 4–6 members. The members of each team sat together and were allowed to communicate and collaborate with each other, as well as to browse the Internet and access the course materials. However, communication or collaboration between different teams was not allowed.

The narrative of the educational escape room began with a video message from an actor playing the Prime Minister of Spain, alerting that a deadly virus has been discovered and that a web application developed by a missing researcher may hold the key to creating a vaccine. The students were charged with the mission of discovering how to use this application based on the existing documentation and to use it for creating the vaccine before the time limit expires. To accomplish the mission, the students were given a classified report containing details about the case and a hyperlink to the vaccine generation application.

The educational escape room consisted of a total of five puzzles arranged in a linear sequence. Therefore, to achieve successful completion of the activity, the students were required to solve the five puzzles in a predetermined order, without the ability to tackle multiple puzzles concurrently. The escape room puzzles combined educational content related to software modeling with game mechanics commonly used in recreational escape rooms. As a consequence, progressing in the activity required the students not only to apply specific knowledge of software modeling, but also to engage in actions characteristic of recreational escape rooms, such as assembling a jigsaw, discovering a concealed symbol, or matching a given pattern.

For example, Fig. 1 shows a puzzle of the educational escape room that required students to correctly interpret UML class diagrams, listen to several voice recordings, and successfully assemble a jigsaw. The topics addressed and the game mechanics involved by each puzzle of the educational escape room are shown in Table I. A detailed description of these puzzles and more details about the narrative can be found in a previous work [14], which reports an earlier experience with the same escape room. In this regard, it is important to note that this previous work [14] did not perform a comparison of the escape room with a traditional approach and did not examine the use of the escape room in face-to-face settings but its use when it is conducted remotely. Therefore, the trial reported in the present article and its contribution are completely new.

With the aim of helping students encountering difficulties completing the educational escape room, an approach for delivering hints based on self-assessed quizzes was employed. The participating teams had the freedom to request hints at their discretion during the activity with the following limitations. First, all hints had to be requested through the Escapp platform. In those cases in which a student requested a hint to the supervising teachers, they instructed the student to request the hint via the Escapp platform. Second, requested hints were only given to a team after passing a five-question quiz on software modeling. Only those quiz attempts in which at least four of the five questions were answered correctly were considered passed. The questions shown in each quiz attempt were randomly chosen from a question bank. Third, teams were obliged to wait for a minimum of 5 min after receiving a hint before they could request another one.

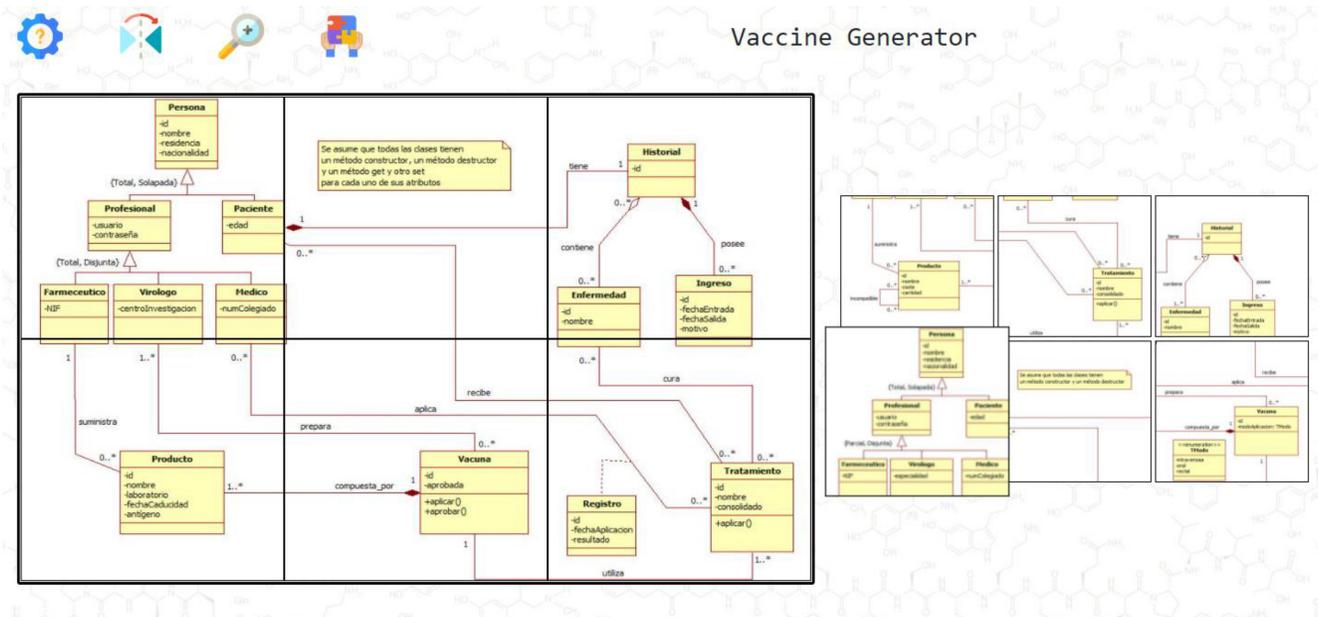

Fig. 1. Escape room puzzle.

TABLE I
ESCAPE ROOM PUZZLES: TOPICS AND GAME MECHANICS

| PUZZLE No. | TOPICS | GAME MECHANICS |
|---|---|---|
| 1 | | - Element of surprise.<br>- Reading.<br>- Searching for objects in images. |
| 2 | - UML use case diagrams.<br>- UML sequence diagrams. | - Pattern identification.<br>- Using something in an unusual way.<br>- Drawing.<br>- Research using information sources. |
| 3 | - UML activity diagrams. | - Searching for hidden objects.<br>- Establishing logical relationships.<br>- Using something in an unusual way. |
| 4 | - UML class diagrams. | - Hearing.<br>- Assembly of an object. |
| 5 | - UML state diagrams.<br>- UML sequence diagrams. | - Noticing something obvious in the environment.<br>- Searching for hidden objects.<br>- Symbol substitution.<br>- Examining images.<br>- Artifact manipulation. |

TABLE II
RESULTS OF THE PRETEST AND THE POST-TEST

| | | CONTROL GROUP (N = 162) | EXPERIMENTAL GROUP (N = 164) |
|---|---|---|---|
| PRE-TEST | M | 6.5 | 6.2 |
| | SD | 1.7 | 1.7 |
| POST-TEST | M | 7.2 | 7.7 |
| | SD | 1.5 | 1.7 |
| LEARNING GAINS | M | 0.7 | 1.5 |
| | SD | 1.3 | 1.7 |
| WILCOXON SIGNED-RANKS TEST FOR PAIRED SAMPLES | p-value | < 0.001 | < 0.001 |
| | Effect size | 0.33 | 0.49 |

Whenever a team requested a hint and successfully completed the triggered quiz, the Escapp platform automatically gave it a hint related to the puzzle it was facing at the moment. Teams could obtain multiple hints per puzzle. The hints were designed in such a way that each hint provided by the Escapp platform for a same puzzle offered students increasingly valuable information compared to the preceding hint. The final hint for each puzzle revealed the solution to that puzzle, guaranteeing that no team would remain stuck indefinitely. Moreover, in some puzzles, teams had the option to choose a specific category for the obtained hint and thereby receive help for a specific aspect or task of the puzzle. More details about the hint approach based on quizzes employed in the escape room and how the Escapp platform supports it can be found in [19].

## III. RESULTS

### A. Learning Gains

Table II shows the post- and pretest scores achieved by the students in each group, along with the within-group differences between these scores (i.e., the learning gains) and their corresponding effect sizes. Statistically significant differences ($p$-value < 0.001) were observed in both groups when comparing the post-test and pretest scores, indicating substantial learning gains. In the control group this difference had a medium effect size ($r = 0.33$), whereas in the experimental group it had a medium to large effect size ($r = 0.49$). These results indicate that the interventions conducted in both groups were effective in terms of knowledge acquisition.

The difference in the pretest scores between groups was not statistically significant and had a less than small effect size, which indicates that the initial knowledge about the covered topics was practically equal in both groups. This was an expected finding given that random sampling was used.

The comparison of the post-test scores between groups, which was performed by using ANCOVA with the pretest scores as covariate, revealed a statistically significant difference with a small to medium effect size ($p$-value < 0.001, $r = 0.25$) in favor of the experimental group. Based on this result, it can be concluded that, in terms of knowledge acquisition, the effectiveness of the educational escape room was higher than that of the traditional lecture.

### B. Students' Perceptions

The results of the questionnaire distributed to collect the perceptions of the students in the experimental group toward the educational escape room are presented in Table III. The overall feedback from the students regarding the learning activity was overwhelmingly positive. The overall opinion of the students on the educational escape room was fantastic and, when they were asked if they would recommend it to other software engineering students or if they would like to find activities of the same type in other courses, practically all students answered in the affirmative. Moreover, the students found the activity fun and immersive. Additionally, they reported that the activity was well organized and they agreed that one and a half hours was an appropriate duration. The students also expressed satisfaction with the manner in which the escape room was overseen, and the majority found the quiz-based approach used to deliver hints to be appropriate.

Regarding self-perceived learning effectiveness, the questionnaire results show that the students perceived an improvement in their knowledge on software modeling as a consequence of playing the educational escape room. This finding is aligned with the learning effectiveness objectively measured through pre- and post-tests. With regard to the comparison of the educational escape room with traditional approaches, the students in the experimental group expressed a strong preference for the escape room over a traditional lecture and most of them agreed that the escape room was more effective than a traditional lecture to learn on software modeling.

With respect to the technology support employed to conduct the educational escape room, it is worth remarking that students had an outstanding opinion on the Escapp platform and found it easy to use. Furthermore, in spite of the educational escape room being conducted face-to-face, the fact that it was based on digital puzzles and was run through the Escapp web platform was very well received by the students. Indeed, students manifested a strong preference for participating in educational escape rooms conducted through the Escapp platform over participating in educational escape rooms without digital support. Overall, the outcomes of the perceptions questionnaire show that the students highly appreciated the utilization of software tools and virtual resources to enhance the educational escape room experience.

TABLE III
RESULTS OF THE PERCEPTIONS QUESTIONNAIRE ($N$=164)

|  | M | SD |
|---|---|---|
| What is your overall opinion on the escape room? (1 Poor – 5 Very Good) | 4.6 | 0.5 |
| What is your overall opinion on the Escapp platform? (1 Poor – 5 Very Good) | 4.6 | 0.6 |
| Please, state your level of agreement with the following statements (1 Strongly disagree – 5 Strongly agree): |  |  |
| The escape room allowed me to improve my knowledge on software modeling. | 3.9 | 0.9 |
| I learned more with the escape room than I would have with a traditional lecture. | 3.6 | 1.1 |
| I liked the escape room more than a traditional lecture. | 4.4 | 0.9 |
| The escape room was fun. | 4.5 | 0.6 |
| The escape room was an immersive experience. | 4.3 | 0.9 |
| The duration of the escape room was adequate. | 4.6 | 0.6 |
| The escape room was well organized. | 4.6 | 0.6 |
| The supervision of the escape room was adequate. | 4.5 | 0.7 |
| The approach used to deliver hints was adequate. | 3.7 | 1.1 |
| I liked the fact that the escape room used digital puzzles. | 4.6 | 0.7 |
| I found the Escapp platform easy to use. | 4.6 | 0.6 |
| I would rather participate in an educational escape room conducted through Escapp than in a similar activity that lacks digital support. | 4.0 | 1.0 |
|  | YES (%) | NO (%) |
| Would you recommend other students to participate in the escape room? | 99.4 | 0.6 |
| Would you like other courses to include activities like the escape room conducted? | 100 | 0 |

## IV. DISCUSSION

The results reported in this article show that the students who learned software modeling through the educational escape room outperformed those who learned through a traditional lecture in terms of knowledge acquisition. These results are consistent with those of the only randomized controlled trial identified by the current systematic literature reviews on educational escape rooms that also compared these activities versus traditional face-to-face teaching, which was conducted by Fuentes-Cabrera et al. [7] and also found that an educational escape room was more effective than a traditional teaching approach. Nevertheless, it should be noted that the escape room analyzed in that prior work [7] was employed to teach mathematics in a secondary education setting and its effectiveness was evaluated through self-reported student perceptions instead of objective data.

The presented results are also consistent with those of the two previous studies that analyzed the learning effectiveness of an educational escape room related to software engineering using a one-group pre-post test design [13], [14]. The effect size of the pre- and post-test score difference was statistically significant and medium in the first of these studies [13] and statistically significant and large in the second one [14].

In this regard, it is worth noting that the effect size in the second of these studies [14] was very similar ($r = 0.53$) to that obtained in this study for the experimental group ($r = 0.49$).

In view of the obtained results, it can also be concluded that the educational escape room had positive impacts on students' perceptions. This finding is consistent with the current body of research on educational escape rooms [2], [3], [4], [5], [6] and, specifically, with those prior research works that examined perceptions of higher-education students toward escape rooms for teaching software engineering topics [9], [11], [13], [14], [15]. Regarding the students' perceptions, it is also worth pointing out that the results of this trial suggest that students have a highly favorable view of virtual educational escape rooms. Specifically, it was found that students showed a high acceptance toward the incorporation of digital puzzles and the use of a web-based platform to carry out the escape room. Other studies that examined students' perceptions toward technology-enhanced educational escape rooms yield consistent findings [19], [20].

As shown by several literature reviews that have analyzed the existing empirical evidence on the impacts of game-based learning [1], [21], [22], [23], [24], [25], [26], a majority of studies comparing game-based instruction and traditional instruction have found that either game-based instruction outperforms traditional instruction in terms of learning effectiveness, or that both approaches had similar learning effectiveness but game-based instruction was preferred by the students. However, some studies in which the learning effectiveness of game-based instruction was found to be worse than that of traditional teaching have also been reported. Regarding this matter, it should be noted that most of the research on game-based learning conducted over the past two decades has focused on the educational application of video games while other types of games have received less attention. For instance, as mentioned earlier, there was little focus on educational escape rooms until 2017. Based on the available evidence, it is reasonable to anticipate that most future studies comparing game-based learning using educational escape rooms versus traditional teaching methods will find evidence in favor of the game-based approach. However, it is also to be expected that in some cases the result will be the opposite. Therefore, future research should examine the factors of educational escape rooms that positively impact game-based learning experiences, just as prior research has done with serious video games [27].

In this regard, it is worth remarking that the present study analyzed the overall learning effectiveness of the educational escape room, but did not analyze the contribution to that effectiveness of individual factors. Therefore, there is no evidence as to which specific features of the educational escape room made the game-based learning experience successful. There are plenty of features that can impact the effectiveness of an educational escape room, such as the use of technology, the type of interaction (face-to-face or at distance), the type of participation (individual or in teams), the narrative, the structure and design of the puzzles, the game mechanics involved, or the approach used to prevent players from becoming stuck. For instance, the analyzed experience could have yielded different results if no software had been used to manage the educational escape room, if virtual reality had been used to increase immersion, if the students had participated remotely instead of in person, if the students had participated individually instead of in teams, if the puzzles had been designed differently, or if free hints had been given instead of applying the quiz-based hint approach previously explained.

In view of these facts, research studies examining escape room features like those mentioned above should be considered highly warranted.

In addition to not analyzing the contribution of specific features of the educational escape room to its overall learning effectiveness, the present study has further limitations. First, the study only tackled one educational escape room designed to teach about a specific topic: software modeling. The puzzles of this escape room blend game mechanics with specific educational content on software modeling and hence they can only be used to teach this topic. Developing an escape room to teach about other topic would require to design and build specific puzzles tailored to address said topic accurately, which should target other learning objectives and could use game mechanics very different from those used by the escape room examined in this work. Therefore, although educational escape rooms share many common characteristics (e.g., narrative, limited time, and puzzles), it should be taken into account that there are several limitations in generalizing the findings of this study to other topics beyond software modeling. Second, the long-term effectiveness of the educational escape room and its benefits to real-world software engineering scenarios were not examined since learning gains were measured only immediately after the interventions. Third, the possible variation in the effectiveness of the interventions depending on the specific characteristics of the participating students was not analyzed. This type of analysis would allow to discover if the educational escape room is more effective for certain student profiles (e.g., for those students who play games more frequently, are more familiarized with game-based learning environments or have stronger teamwork and communication skills), as well as to analyze the influence of variables such as age and gender.

In summary, although it is undeniable that the randomized controlled trial presented in this article provides solid and novel empirical evidence on the learning effectiveness of educational escape rooms, it is still not enough to draw conclusive claims on this matter. Overall, the existing evidence on educational escape rooms suggests that these activities, when properly designed and conducted, can positively impact students' learning and may even be more effective than traditional teaching methods. However, further empirical studies with robust research designs are needed to confirm these findings, examine to what extent they can be generalized to all educational settings and disciplines, understand the impacts of educational escape rooms in the long term and on specific student profiles, and determine how these novel learning activities should be designed and run to be effective.

## V. Conclusion

This article provides strong and novel empirical evidence that game-based learning using educational escape rooms can be more effective than traditional teaching in improving students' knowledge on software modeling, a core topic in software engineering education. Although previous works reported on the use of educational escape rooms in higher-education settings for teaching topics closely related to software engineering [8], [9], [10], [11], [12], [13], [14], [15], none of them analyzed the learning effectiveness of the interventions through a control and experimental group design. Thus, this article makes a novel and significant contribution by reporting, for the first time, a comparison of the learning effectiveness of educational escape rooms and traditional teaching in software engineering education by means of a randomized controlled trial.

The findings of this article provide educators with valuable insights into the benefits of using educational escape rooms that they can use to make informed decisions about incorporating these novel learning activities into their teaching practice. In order to help educators interested in using escape rooms as teaching tools, some practical guidelines for designing and conducting educational escape rooms based on the lessons learned are suggested below.

1) Encourage students to participate in the educational escape room as part of a team in order to provide them with the opportunity of developing soft skills, such as communication, teamwork, and leadership.
2) Design the educational escape room to be completed in 1–2 h. Shorter durations could prevent students from mastering meaningful learning objectives, while longer durations would increase the risk of students becoming tired or losing interest.
3) Provide participating students with clear initial guidance outlining the rules of the educational escape room, its final goal, the initial steps to take, and the mechanism available to request help if they encounter doubts or become stuck on a certain puzzle.
4) Employ a linear puzzle structure so that each escape room puzzle cannot be completed until the previous ones have been solved. Organizing puzzles sequentially ensures that all team members face all puzzles (and thus engage with all educational content) and makes it easier to monitor the educational escape room and to provide timely help to those students who needed it.
5) Connect all escape room puzzles through an engaging narrative and ensure that every puzzle, task, and resource presented to the students had a reason consistent with the narrative to exist. As a result, the educational escape room will have a coherent story that is more likely to motivate students.
6) Design the escape room puzzles intertwining educational content with various game mechanics and in such a way that the probability of students solving them through trial and error is minimal. This approach allows to achieve an appropriate balance between the educational and playful aspects.
7) Increase the difficulty of the escape room puzzles gradually and continuously. Starting with a short and easy-to-solve puzzle makes it easier for students to familiarize themselves with the environment and complete the first steps of the educational escape room without the risk of becoming discouraged.
8) Conduct educational escape rooms by using software tools designed specifically for this purpose like the Escapp platform [19]. The use of this kind of tools can greatly ease various activities needed to conduct educational escape rooms, such as enrollment, content delivery, checking of puzzle solutions, monitoring, and hint management. In this regard, it is worth indicating that the students surveyed in this study had very positive perceptions toward the Escapp platform and expressed a strong preference for conducting educational escape rooms using this kind of software.
9) Deliver hints using an approach like the one described in Section II-F, which requires students to pass self-assessed quizzes in order to obtain hints. This quiz-based hint approach provides a great opportunity to cover concepts that are complex to address through the escape room puzzles, as well as to adjust the difficulty of the educational escape room.

Although this article shows that educational escape rooms can be very useful for teaching software modeling, it is not clear whether its findings are generalizable to other topics. Therefore, future research should analyze the effectiveness of these activities to teach other topics (e.g., software engineering topics such as software development life cycle, software design, software quality, and software verification and validation) through well-designed experimental studies, such as randomized controlled trials. Moreover, given that long-term effectiveness of educational escape rooms has not been previously examined, longitudinal studies should be conducted to provide insights on the sustainability of the learning outcomes. Future research could also compare the effectiveness between educational escape rooms and other learning alternatives, such as intelligent tutoring systems, serious video games, project-based learning, and video-based learning. Thereby, educators could be provided with valuable insights that would help them to choose the most suitable teaching method depending on the topic to be addressed and the characteristics of the educational setting and the learners.

In this regard, it is worth pointing out that the future studies analyzing the impacts of educational escape rooms on specific learner profiles will also help educators making these decisions. Finally, it would also be very interesting to conduct studies that contribute to a better understanding of how to design and run effective educational escape rooms, such as experiments comparing educational escape rooms experiences that differ in only one characteristic. This kind of experiments would enable to analyze the impact of specific features of educational escape rooms on their effectiveness and thereby provide insights that allow teachers to implement better game-based learning experiences using educational escape rooms.

## REFERENCES

[1] C. Udeozor, R. Toyoda, F. Russo Abegão, and J. Glassey, "Digital games in engineering education: systematic review and future trends," *Eur. J. Eng. Educ.*, vol. 48, no. 2, pp. 312–329, 2022, doi: 10.1080/03043797.2022.2093168.

[2] P. Fotaris and T. Mastoras, "Escape rooms for learning: A systematic review," in *Proc. 13th Eur. Conf. Games Based Learn. (ECGBL)*, 2019, pp. 1–9, doi: 10.34190/GBL.19.179.

[3] A. Veldkamp, L. van de Grint, M. C. Knippels, and W. van Joolingen, "Escape education: A systematic review on escape rooms in education," *Educ. Res. Rev.*, vol. 31, Nov. 2020, Art. no. 100364, doi: 10.1016/j.edurev.2020.100364.

[4] A. Makri, D. Vlachopoulos, and R. Martina, "Digital escape rooms as innovative pedagogical tools in education: A systematic literature review," *Sustainability*, vol. 13, no. 8, p. 4587, 2021, doi: 10.3390/su13084587.

[5] C. Lathwesen and N. Belova, "Escape rooms in STEM teaching and learning—Prospective field or declining trend? A literature review," *Educ. Sci.*, vol. 11, no. 6, p. 308, 2021, doi: 10.3390/educsci11060308.

[6] L. H. Taraldsen, F. O. Haara, M. S. Lysne, P. R. Jensen, and E. S. Jenssen, "A review on use of escape rooms in education–touching the void," *Educ. Inq.*, vol. 13, no. 2, pp. 169–184, 2022, doi: 10.1080/20004508.2020.1860284.

[7] A. Fuentes-Cabrera, M. E. Parra-González, J. López-Belmonte, and A. Segura-Robles, "Learning mathematics with emerging methodologies—The escape room as a case study," *Mathematics*, vol. 8, no. 9, p. 1586, 2020, doi: 10.3390/math8091586.

[8] C. Borrego, C. Fernández, I. Blanes, and S. Robles, "Room escape at class: Escape games activities to facilitate the motivation and learning in computer science," *J. Technol. Sci. Educ.*, vol. 7, no. 2, pp. 162–171, 2017, doi: 10.3926/jotse.247.

[9] A. M. Ho, "Unlocking ideas: Using escape room puzzles in a cryptography classroom," *Primus*, vol. 28, no. 9, pp. 835–847, 2018, doi: 10.1080/10511970.2018.1453568.

[10] S. Seebauer, S. Jahn, and J. Mottok, "Learning from escape rooms? A study design concept measuring the effect of a cryptography educational escape room," in *Proc. IEEE Global Eng. Educ. Conf. (EDUCON)*, pp. 1684–1685, 2020, doi: 10.1109/EDUCON45650.2020.9125333.

[11] K. R. Otemaier, E. E. Grein, P. G. Zanese, and N. S. Bosso, "Educational escape room for teaching mathematical logic in computer courses," in *Proc. SBGames*, 2020, pp. 595–604.

[12] J. Saltz and R. Heckman, "Using structured pair activities in a distributed online breakout room," *Online Learn. J.*, vol. 24, no. 1, pp. 227–244, 2020, doi: 10.24059/olj.v24i1.1632.

[13] S. López-Pernas, A. Gordillo, E. Barra, and J. Quemada, "Analyzing learning effectiveness and students' perceptions of an educational escape room in a programming course in higher education," *IEEE Access*, vol. 7, pp. 184221–184234, 2019, doi: 10.1109/ACCESS.2019.2960312.

[14] A. Gordillo, D. López-Fernández, S. López-Pernas, and J. Quemada, "Evaluating an educational escape room conducted remotely for teaching software engineering," *IEEE Access*, vol. 8, pp. 225032–225051, 2020, doi: 10.1109/ACCESS.2020.3044380.

[15] S. López-Pernas, A. Gordillo, E. Barra, and J. Quemada, "Examining the use of an educational escape room for teaching programming in a higher education setting," *IEEE Access*, vol. 7, pp. 31723–31737, 2019, doi: 10.1109/ACCESS.2019.2902976.

[16] M. Ardis, D. Budgen, G. W. Hislop, J. Offutt, M. Sebern, and W. Visser, "Curriculum guidelines for undergraduate degree programs in software engineering," *Computer*, vol. 48, no. 11, pp. 106–109, Nov. 2015.

[17] L. J. Cronbach, "Coefficient alpha and the internal structure of tests," *Psychometrika*, vol. 16, no. 3, pp. 297–334, 1951, doi: 10.1007/BF02310555.

[18] J. Cohen, *Statistical Power Analysis for the Behavioral Sciences*, 2nd ed. New York, NY, USA: Routledge, 1988.

[19] S. Lopez-Pernas, A. Gordillo, E. Barra, and J. Quemada, "Escapp: A web platform for conducting educational escape rooms," *IEEE Access*, vol. 9, pp. 38062–38077, 2021, doi: 10.1109/ACCESS.2021.3063711.

[20] M. Monnot, S. Laborie, G. Hébrard, and N. Dietrich, "New approaches to adapt escape game activities to large audience in chemical engineering: Numeric supports and students' participation," *Educ. Chem. Eng.*, vol. 32, pp. 50–58, Jul. 2020, doi: 10.1016/j.ece.2020.05.007.

[21] M. H. Hussein, S. H. Ow, L. S. Cheong, M.-K. Thong, and N. Ale Ebrahim, "Effects of digital game-based learning on elementary science learning: A systematic review," *IEEE Access*, vol. 7, pp. 62465–62478, 2019, doi: 10.1109/ACCESS.2019.2916324.

[22] U. Tokac, E. Novak, and C. G. Thompson, "Effects of game-based learning on students' mathematics achievement: A meta-analysis," *J. Comput. Assist. Learn.*, vol. 35, no. 3, pp. 407–420, 2019, doi: 10.1111/jcal.12347.

[23] T. Hainey, T. M. Connolly, E. A. Boyle, A. Wilson, and A. Razak, "A systematic literature review of games-based learning empirical evidence in primary education," *Comput. Educ.*, vol. 102, pp. 202–223, Nov. 2016, doi: 10.1016/j.compedu.2016.09.001.

[24] C. A. Bodnar, D. Anastasio, J. A. Enszer, and D. D. Burkey, "Engineers at play: Games as teaching tools for undergraduate engineering students," *J. Eng. Educ.*, vol. 105, no. 1, pp. 147–200, 2016, doi: 10.1002/jee.20106.

[25] E. A. Boyle et al., "An update to the systematic literature review of empirical evidence of the impacts and outcomes of computer games and serious games," *Comput. Educ.*, vol. 94, pp. 178–192, Mar. 2016, doi: 10.1016/j.compedu.2015.11.003.

[26] T. M. Connolly, E. A. Boyle, E. MacArthur, T. Hainey, and J. M. Boyle, "A systematic literature review of empirical evidence on computer games and serious games," *Comput. Educ.*, vol. 59, no. 2, pp. 661–686, 2012, doi: 10.1016/j.compedu.2012.03.004.

[27] W. S. Ravyse, A. S. Blignaut, V. Leendertz, and A. Woolner, "Success factors for serious games to enhance learning: a systematic review," *Virtual Real.*, vol. 21, pp. 31–58, Mar. 2017, doi: 10.1007/s10055-016-0298-4.